# Open Access Books in the Humanities and Social Sciences: an Open Access Altmetric Advantage.

Michael Taylor, University of Wolverhampton; Digital Science

## Abstract

The last decade has seen two significant phenomena emerge in research communication: the rise of open access (OA) publishing, and evidence of online sharing in the form of altmetrics. There has been limited examination of the effect of OA on online sharing for journal articles, and little for books. This paper examines the altmetrics of a set of 32,222 books (of which 5% are OA) and a set of 220,527 chapters (of which 7% are OA) indexed by the scholarly database Dimensions in the Social Sciences and Humanities. Both OA books and chapters have significantly higher use on social networks, higher coverage in the mass media and blogs, and evidence of higher rates of social impact in policy documents. OA chapters have higher rates of coverage on Wikipedia than their non-OA equivalents, and are more likely to be shared on Mendeley. Even within the Humanities and Social Sciences, disciplinary differences in altmetric activity are evident. The effect is confirmed for chapters, although sampling issues prevent the strong conclusion that OA facilitates extra attention at whole book level, the apparent OA altmetrics advantage suggests that the move towards OA is increasing social sharing and broader impact.

## Introduction

Two of the largest phenomena in scientific communication in the last decade have been the rise of Open Access (OA) journal publications, and the research area known as altmetrics. OA publications are those that are freely available on the internet, via a range of routes. Altmetrics is the collection, reporting and analysis of attention being paid to research publications across a variety of online platforms.

Books and book chapters are under-represented in the growing corpus of research on OA. It has been suggested that this has arisen due to the general lack of attention paid to both the book form, and also the Arts and Humanities disciplines – which tend to favour books as their preferred channel for publishing research.

In general, the absence of sales figures, reliable metadata, download figures and the relatively slow citation performance of books has made a comparative study of OA versus non-OA books and chapters challenging. The relative paucity of data may have contributed to a low level of examination in the scientometric literature, and as such, there is a lack of compelling evidence to drive the adoption of OA for books and chapters.

An emerging literature examining the effect on social sharing and broader impact on OA journal articles offers some methodological insights, but given the known differences between journal articles and books, do not offer any results that may be extrapolated to books.

This research uses two new data sources (Unpaywall, Dimensions) that contain more data about books and chapters that have been readily available before, in order to analyse the extent to which OA and non-OA books and chapters might differ in social sharing and broader impact, as reported by the Altmetric.com service.

## The growth of Open Access

The OA movement began in 2001 (Suber, 2012), and over the years, several different classes of OA publishing have emerged, usually referred to by colours:

- Gold OA – or the so-called 'pay to publish' model – occurs when a publisher receives a fee for making the content available for free on the journal's website, along with an associated license that allows for reused.

- Hybrid OA applies to paid-access journals that publish articles OA if the author pays an optional OA fee.

- Green OA applies when an article is saved in an OA repository:
    - 'Published' Green refers to the saving of the *final, published* article.
    - 'Accepted' Green refers to the saving of the *accepted* article.
    - 'Submitted' Green refers to the saving of the *submitted* article.

- Bronze OA is a relatively new term, defined as a document that is freely on a publisher's website, but without any license being made available (Piwowar et al., 2018a).

- Black OA refers to the unauthorised sharing of research output, e.g., on SciHub. Black OA is excluded from this research.

There has been a sustained growth in the rates of OA article publishing, from a reported rate of 20.4% in 2009 (Björk et al., 2010), to 45% in 2015 (Piwowar et al., 2018b). There are significant disciplinary differences, with an OA rate of 50% being reported in the biomedical sciences in 2010 (Kurata, Morioka, Yokoi, & Matsubayashi, 2013). This trend was later confirmed, with Biomedical and Physical Sciences having the highest rates of OA, and the Social Sciences and Engineering having the lowest rates (Piwowar et al., 2018b).

The OA model has not been so influential for scholarly books, with only 7165 OA books being reported by Dimensions for 2013[1], from an estimated total 86,000 monographs being published that year (Grimme et al., 2019). That current models for funding OA were not well suited for books was recognized in 2012 (Pinter & Thatcher, 2012). The same concerns were still being addressed by Grimm et al, half a decade later.

The growth in OA seems to have occurred primarily because of mandates and policies. Since 2006, there have been a succession of initiatives to implement OA mandates most notably in Canada, the Europe Union, and the USA in 2009 (Canadian Institutes of Health

---

1   https://app.dimensions.ai/discover/publication?or_facet_publication_type=monograph&or_facet_publication_type=book&or_facet_year=2013&or_facet_open_access_status_free=oa_all December 2019

Research, 2006; European Commission, 2008; National Institutes of Health, 2009). Plan S is a later initiative (Schiltz, 2018), organized by a significant group of governments, funders and other institutions, with the stated ambition of making research publications funded or supported by them OA by 2021 (Coalition S, 2018).

Policies aimed at increasing the rate of OA publishing in journals – such as Plan S – have been criticized for not taking into consideration certain issues of particular concern to the Arts, Humanities and Social Sciences (Frantsvåg & Strømme, 2019), and – in particular – that the focus of Plan S fails to address the need to drive public engagement and support the 'democratization of research' (Holbrook, 2019). Plan S has also been criticized for failing to adequately address the issue of OA publishing for books (Science Europe, 2019).

The size of the academic book market is considered to be stable although the rate of metadata deposition with Crossref is increasing (Grimme et al, 2019).

Many academic publishers support OA for books and chapters with both Green and Gold options. Self-archiving in Green repositories is commonly permitted, often featuring an embargo period of between 12-24 months (OAPEN, 2020). Taylor and Francis, Springer-Nature, OUP and several small presses are signatories to the OAPEN list of compliant publishers. Wiley supports Green self-archiving (Wiley, 2020). Elsevier – in common with most book publishers – offers a Gold/Hybrid route, but unlike other major scholarly book publishers, have no published policy on Green self-archiving (Elsevier, 2020b).

### Altmetrics and OA

The term 'altmetrics' was introduced in 2010 (Priem, Taraborelli, Groth, & Neylon, 2010) to bring together a number of discrete and disparate social web attention sources under one umbrella to "reflect the broad, rapid impact of scholarship". Many of the areas of focus contained within altmetrics had already been studied for over the preceding decade, under the name of 'webometrics' (Almind & Ingwersen, 1997), which, in itself had emerged from the field of bibliometrics and scientometrics. Pioneers in this emergent field had been explicitly analysing web-traffic, usage and content to understand the emerging online world in the context of research evaluation and scientometrics since the late-90s (Bar-Ilan, 2000; Thelwall, 2000). The initially distinctive feature of altmetrics was its focus on social web services with an applications programming interface (API), although the term now also encompasses traditional webometrics.

The field is supported by four providers of altmetrics data. Altmetric.com and Plumx were both launched in 2011, followed by Crossref Event Data in 2016 and Cobalt Metrics in 2018[2]. Altmetrics have been collected for many parts of the web, including Wikipedia, news and broadcast media, blogging platforms, social and scholarly networks (Thelwall, Haustein, Larivière, & Sugimoto, 2013), policies (McLeish, 2016) and patents (Altmetric, 2018). Altmetrics have been studied with a view to understanding future citation rates (Eysenbach, 2011), non-scholarly usage (Mohammadi, Thelwall, Haustein, & Larivière, 2015) and social impact (Bornmann, 2014).

It has been known since the early 1990s that media coverage can influence academic behaviour: a study on the citation effect of coverage in the *New York Times* during a year-long strike revealed that mass media coverage effectively "amplified the transmission of

---
2

research" (Phillips, Kanter, Bednarczyk, & Tastard, 1991). This seems likely to extend to media distributed and promoted online.

Although altmetrics might reflect public engagement with research, they might also reflect communication within academia that is merely happening within a public arena (Sugimoto & Larivière, 2017). For example, about half of the tweets mentioning journal articles are from academics (Mohammadi, Thelwall, Kwasny, & Holmes, 2018), despite them forming a small minority of social web users. Policy makers and funders have increasingly required researchers to plan for broader impact (Bornmann, 2013), requiring them to hone their impact management plans (Britt Holbrook & Frodeman, 2011), and this may include promoting their own work online.

Tracing social impact is complex, and requires more dynamic solutions than usual bibliometric approaches (Holmberg, Bowman, Bowman, Didegah, & Kortelainen, 2019). Nevertheless, researchers have found evidence of social impact in social media (Pulido, Redondo-Sama, Sordé-Martí, & Flecha, 2018), and have discussed the importance of positioning argument correctly in policy documents to optimize social impact (Williams, 2018). The use of Wikipedia as a medium to inform patients (Heilman et al., 2011) and respond to their concerns (Didegah, Ghaseminik, & Alperin, 2018) has been investigated.

If OA increases the value of research for academics or the broader public then this increased activity should be detected by comparing the altmetrics of OA and non-OA publications.

The phenomena of a potential increased rate of citation for OA research outputs is known as an OA Citation Advantage (OACA), so the term *OA Altmetrics Advantage* (OAAA) is used here to describe this concept.

Early research into OACA found mixed results, with no significant effect reported in the fields of dermatology (Umstattd, Banks, Ellis, & Dellavalle, 2008) and astrophysics (Kurtz & Henneken, 2007), with the latter arguing that the apparent advantage was due to the early availability of documents in Arxiv. While this observation was potentially confirmed for the first year of an article's life, the advantage was observed to disappear over subsequent years (Davis, Lewenstein, Simon, Booth, & Connolly, 2008).

Subsequent research confirmed the existence of a persistent OACA for most types of OA article, the exception being Gold, where an early OACA disappears (Piwowar et al., 2018b). An important methodological issue is that it is difficult to prove cause-and-effect. All journals are unique, so it is impossible to have a controlled experiment comparing Gold and non-Gold OA journals. For Green OA, if OA articles are more cited than non-OA articles, this could be because researchers are more likely to post their own articles online if they believe them to be important. Thus, a simple comparison of citation rates does not allow a conclusion that OA causes additional citations.

The presence of an OAAA for the volume of attention on both Twitter and Mendeley for a number of articles in a single hybrid journal has been reported (Adie, 2014), who additionally identified an absence of an OAAA for blogs and news sources. These negative finds would have been expected, as the research used mean and median values, an approach not well-suited to the analysis of low-frequency indicators, such as news and blogs.

4 Open Access Books in the Humanities and Social Sciences

The presence of an OAAA for Wikipedia has been reported at a journal-level (Teplitskiy, Lu, & Duede, 2017) – albeit as a secondary factor to the journals' academic status. This finding was potentially weakened, as it only considered the OA status at a journal level, and many journal articles are made OA at an article level, for example through funder mandates and researcher self-archiving.

A study of Finnish papers confirmed the existence of this phenomena for certain fields and attention sources, but a *disadvantage* for other fields and attention sources (Holmberg, Hedman, Bowman, Didegah, & Laakso, 2019). This research focussed largely on the most populous altmetric indicators (Twitter and Mendeley), plus citations from the Web of Science. Other altmetric indicators (news, blogs, Wikipedia and Facebook) were compounded. This research used the OA journal-status, as defined by the Directory of Open Access Journals (DOAJ), meaning that OA articles in hybrid journals, and Green OA articles would have been treated as non-OA. Three social sciences were analysed at a journal-level, with both Psychology, Educational Sciences and Social and Economic Geography showing an OAAA for the compounded indicator. Mendeley showed a negative effective for Psychology.

The existence of strong disciplinary differences in altmetric data for the all journal articles had been reported for all major attention sources, but without focussing on OA status, with links to Digital Humanities research being shared more often on Twitter than Economics (Holmberg & Thelwall, 2014). Psychology was consistently amongst the Scopus subjects with the highest coverage for mass media (3.5%), social networks (36.1%), scholarly networks (71.4%) and a combined indicator for blogs and post-publication peer-review (4.1%), and Economics, Social Sciences and Arts & Humanities being significantly lower for mass media, (0.97-1.56%), social networks (15.2-20.2%), scholarly networks (52.5-63.7%) and the combined indicator (1.8-2.2%) (Taylor, 2015).

## The impact of the scholarly book

Understanding the impact of books has been hindered in several ways. Both book and chapter citations behave differently, both from each other and from journal articles (Chi, 2016). Usage is more heterogeneous and is potentially harder to capture, and hence is less well covered by the tools used in mainstream scientometric analysis (Halevi, Nicolas, & Bar-Ilan, 2016). Nevertheless, there have been attempts to increase the detection and reporting of book-specific impact by Springer (Hawkins, 2016), Altmetric (Torres-Salinas, Gorraiz, & Robinson-Garcia, 2018) and PlumX (Torres-Salinas, Robinson-Garcia, & Gorraiz, 2017). Additional sources have been in investigated, and the disproportionate importance of books to the Arts, Social Sciences and Humanities has been reported, with the caveat that only a relatively small proportion of scholarly books are available for analysis in the major abstract and index databases (Kousha & Thelwall, 2015).

The metadata infrastructure for books offers a significant challenge to researchers. The almost universal ISBN system does not support the free and open distribution of metadata in a manner analogous to Crossref and Datacite (O'Leary & Hawkins, 2019). The likely disproportionate prevalence of Digital Object Identifier (DOI) usage for book chapters by OA publishers influences DOI-based altmetrics and citations gathered for books, making comparisons between OA and non-OA difficult. This may have been the cause of an OACA for book chapters in Conservation Biology Calver & Bradley (2009), for example.

5 Open Access Books in the Humanities and Social Sciences

Few studies have investigated OACA or OAAA for books, perhaps as a result of the lack of systematically available metadata. Snijder, (2016) found a slight OACA and OAAA for Twitter for OA books over a five-year time span for 400 monographs, of which 271 were OA and 129 non-OA. The researchers, noting that since this corpus was only identifiable using ISBN, and that (at that time) ISBNs weren't being tracked by Altmetric.com, were obliged to use a combined heuristical and manual manual approach to identify tweets to the books in their dataset.

In an exploratory paper Wennström et al., (2019) studied a very limited number of OA books (N = 22), highlighting disciplinary differences across all metrics for open monographs, the potential for altmetric data, and the need for more research into both the metrics and author attitudes towards book metrics.

This paper attempts to address some of the deficiencies in the literature to date, by analysing a very large set of books and chapters across a range of both low-frequency and high-frequency attention sources, and to take into account all forms of OA publishing.

# Objectives

The absence of any prior systematic research into OAAA for books is an important omission given their importance in the arts, humanities and many social sciences. The following research questions address this gap:

1. Are OA books and chapters more likely to received attention from News, blogs, Wikipedia, Twitter, Mendeley and Policy attention sources?

2. Is there an OAAA for OA books and chapters, when considering the number of tweets and Mendeley readers?

3. Is the rate of OA publishing increasing for books and chapters?

4. How significant are variations in OAAA between the disciplines that make up the arts, humanities and social sciences?

5. To what extent does the OAAA for books and chapters differ, both from each other and from the journal OAAA?

# Methods

The research design was to gather a large sample of books and book chapters with altmetric records and to compare the altmetric data between the OA and non-OA subsets.

## Data

Digital Science's Dimensions platform (Hook, Porter, & Herzog, 2018) was used as the book source, because it has indexed over 1M monographs and edited volumes, and over 9M chapters, making it the largest index of its type. In contrast, Clarivate's Book Citation Index contains 60,000 books (Clarivate, 2020) and Elsevier's Scopus contains 120,000 (Elsevier, 2020a).

All monographs and edited books (collectively referred to as books), and individually indexed chapters that had been assigned (Herzog, Sorensen, & Taylor, 2016) into the Dimensions

Fields of Research[3] categories covering Arts, Humanities and Social Sciences were extracted as the initial sample. Some are OA, and some are non-OA.

Dimensions' content process starts with harvesting metadata from Crossref, Pubmed and Pubmed Central. In order to apply a Field of Research category, the full text needs to be available: in 2019, Digital Science reported that over 100 of the largest scholarly publishers had supplied full-text for indexing, permitting classification of over two-thirds of the entire data. Dimensions does not select for *inclusion*, rather it follows community-led *exclusion* recommendations (Bode, Herzog, Hood, & McGrath, 2019).

Records for 32,222 books and 204,538 chapters were retrieved (Taylor, 2020), fitting the criteria of:

1. Published between 2013 and 2016,

2. Providing a minimum of 5000 data points, and

3. Having been assigned the categories of Studies in Human Society (FoR code 16), Psychology and Cognitive Sciences (17), Philosophy and Religious Studies (22), Law and Legal Studies (18), Language, Communication and Culture (20), Education (13), Economics (14), and Commerce, Management, Tourism and Services (15). Disciplines that contained fewer than 10 OA books were discarded from the analysis, e.g. Planning and Creative Arts.

Several classifications for OA status are provided by Dimensions: however for the purpose of this research, the various 'OA' indicators (Gold, Hybrid, Green - submitted, published and accepted - and Bronze) were treated as a common OA indicator. Dimensions uses data from [Unpaywall](), the most comprehensive database of OA indicator available (Piwowar et al., 2018b), to classify the status of its books and chapters [4]. The eight disciplines have significant populations of books and chapters (Appendix Table 1), with at least 1619 books and 14,919 chapters each. The percentages vary from 2% to 14%, with the smallest OA number in any category being 11.

Data from Altmetric.com was incorporated from a static dataset provided by Altmetric under a research license, with data covering up to October 2019. Mendeley readership counts were accessed during October 19-20, 2019, using DOI searches in the Mendeley API.

To enable benchmarking of book and chapter performance against journal articles, data from an equivalent set of journal articles – matching subject area, publishing date and OA status at article level – were accessed from Dimensions.

## Analysis

To make comparisons between OA and non-OA books and OA and non-OA chapters, and to enable a comparison with previously reported trends, coverage indicators were calculated for all attention sources, and an average value for Mendeley and Twitter. To answer the research questions, these values were calculated across time, by each subject area, and for books and chapters, each group being divided by OA status.

---

3   https://www.abs.gov.au/AUSSTATS/abs@.nsf/Lookup/1297.0Main+Features12008

4 'Hybrid' books, i.e. Gold books in otherwise non-OA book series were treated as Gold.

The proportion of books or chapters with non-zero attention was analysed because, with the exception of Mendeley and Twitter, the great majority of data points associated with books and chapters are zero. In particular, Policy documents, Wikipedia, News and Blogs typically are reported for fewer than 2% for chapters (Table 3) and fewer than 5% for books. Values based on geometric means were calculated for Twitter (unique accounts) and Mendeley readers since these sources were usually non-zero (see paragraph below for details). OAAAs were then estimated by dividing either the proportion or average for a particular OA subset by the global proportion or average. This generates a normalised Attention Factor (AF) that quantifies any OA advantage (above 1.00) or disadvantage (1.00).

Extreme, outlying values for citations and altmetrics can skew values based on arithmetic means (Hammarfelt, 2014; Ottaviani, 2016; Thelwall, 2017); to minimise this phenomena, this paper focusses on the proportions of the populations that have any altmetric activity for the six attention sources, and reports only average values for the two most populous indicators (i.e., Mendeley and Twitter); using values based on a geometric mean (Thelwall & Fairclough, 2015) rather than the common arithmetic mean. Accordingly, all Twitter and Mendeley values were incremented by 1, and the natural log calculated. These are then averaged, with the exponential of the total being calculated, and decreased by 1. The effect of averaging the natural log is to decrease to influence of any extreme outlying values.

Fisher Exact 2x2 tests – a test optimized for non-parametric and unequal set of populations - were used to calculate the statistical significance of coverage. This tests whether there is a significant difference between the expected and the observed frequencies with populations with one or more categories. A statistically significant result gives evidence that the OA scores tend to differ, on average, to the non-OA scores. Two sample t-tests applied to the logged values were used to assess whether the rates for the OA and non-OA sets were statistically significantly different.

# Results

## Growth in OA books and chapters

Data retrieved from Dimensions (Table 1) suggests that there is no evidence that there is an increase in OA books in all fields between 2013 and 2015, with a rise indicated between 2015-2016. In contrast, the volume of OA chapters appears to increase across the sampling period.

Table 1 Absolute volume of books and chapters published in all FoR codes ([Dimensions,](Dimensions,) retrieved October 22, 2019)

| Publication year | 2013 | | 2014 | | 2015 | | 2016 | |
|---|---|---|---|---|---|---|---|---|
| | OA | non-OA | OA | non-OA | OA | non-OA | OA | non-OA |
| Books | 379 | 9107 | 324 | 6489 | 349 | 4939 | 480 | 10155 |
| Chapters | 3260 | 46476 | 3333 | 44664 | 4043 | 58361 | 5353 | 55037 |

The eight disciplines have significant populations of books and chapters (Appendix Table 1). The two most populous are 'Studies in Human Society' which contains 9675 books and 54,440 chapters (of which 4.56% and 6.78% respectively are OA) and 'Language, Communication and Culture' with 7915 books, of which 4.26% are OA, and 'Psychology and Cognitive Sciences', with 39,477 chapters (of which 7.61% are OA).

The two smallest disciplines for books are 'Commerce, Management, Tourism and Services' with 1619 books (5.13% are OA) and 'Economics' with 1909 (7.02% are OA). The two smallest disciplines for chapters are 'Law and Legal Studies', with 15,237 (of which 6.59% are OA) and 'Philosophy and Religious Studies', with 14,919 (of which 5.03% are OA).

Chapters dominate the population of all subject areas, with substantial disciplinary differences. They range from 77.4% ('Language, Communication and Culture') to 93.7% ('Commerce, Management, Tourism and Services') (Appendix Table 1). There are no discernable longitudinal trends (Appendix Table 2).

The rate of OA publishing for books and chapters in the eight Humanities and Social Sciences fields covered by this paper is considerably lower than an equivalent set of journal articles (Figure 1) for all subject areas. Only the fields of Economics, and 'Commerce, Business and Management' a comprable proportion to journal articles, when they became the only two fields to achieve over 10% OA for books and chapters in 2016. In contrast, the proportion of OA journal articles in these fields range from 22.8% to 41.9% in 2013 to 47.6% and 57.1% in 2016 (Philosophy, and Economics, respectively).

The distributions of OA type vary by published format, with books and chapters showing low levels of Gold publishing, and high levels of Green, Submitted, when compared with an equivalent set of journal articles (Figure 2).

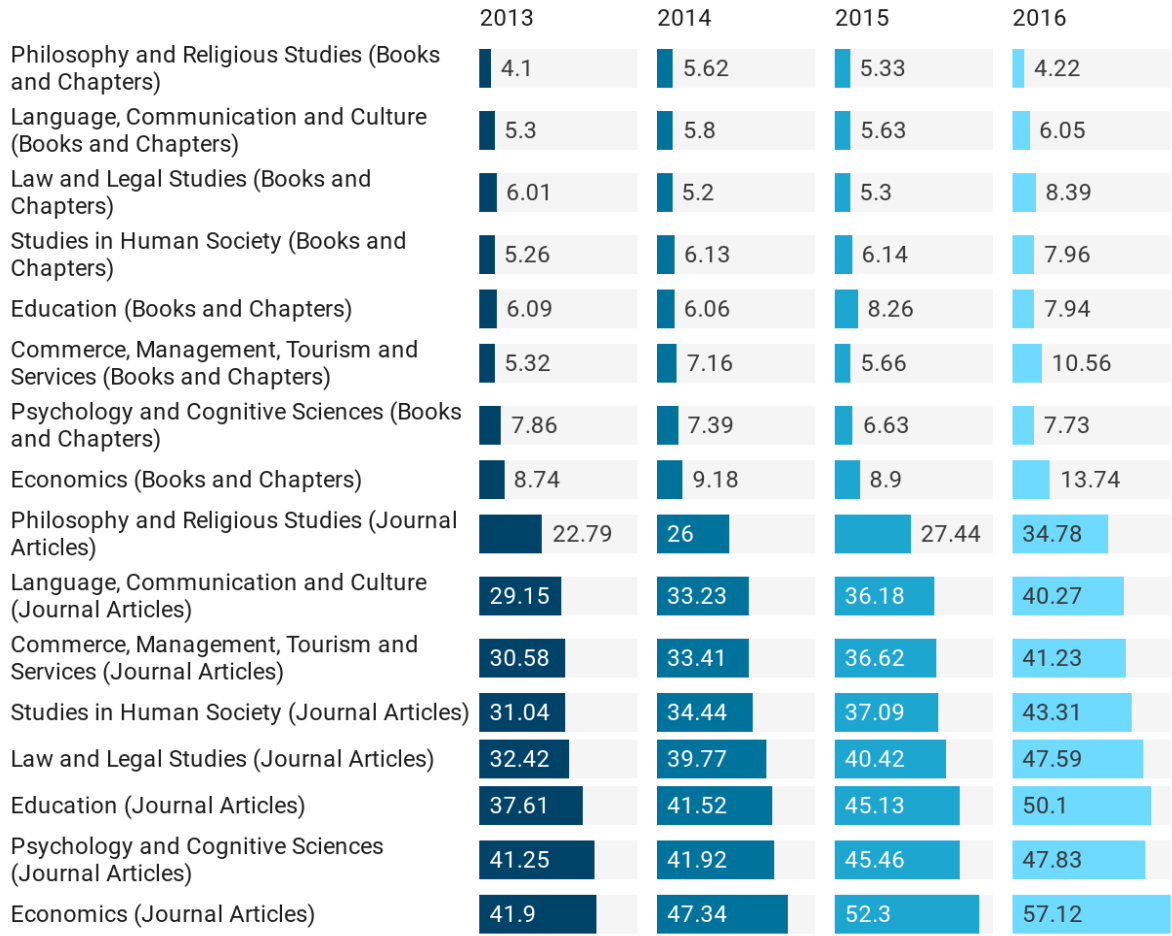

*Figure 1 Relative Growth in OA for Journal Articles and Books and Chapters (combined) for Social Sciences and Humanities (%). (Dimensions, retrieved December 19, 2019).*

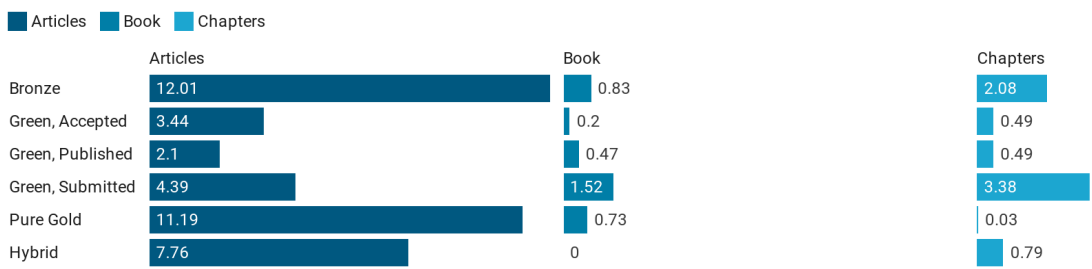

*Figure 2 Distribution of OA Types by Publication Format (%). (Dimensions, retrieved December 19, 2019).*

## General trends in the altmetrics of books and chapters

The proportion of books with any altmetric attention (as defined by having a minimum of one Tweet, Blog, News, Wikipedia or Policy Citation, or one Mendeley reader) remains stable over the four years (Table 2), with no OAAA apparent for books. The proportion of OA chapters with any altmetric attention is consistently higher for all years, showing an overall OAAA over non-OA chapters of 16.7% in 2013 and 12.6% in 2016.

*Table 2 Proportion of Books and Chapters with Altmetrics Attention Over Time (Sources: Altmetric and Mendeley)*

| Publication year | Books | | Chapters | |
| --- | --- | --- | --- | --- |
| | OA | non-OA | OA | non-OA |
| 2013 | 65.70% | 65.97% | 68.65% | 51.99% |
| 2014 | 66.05% | 60.59% | 69.58% | 54.81% |
| 2015 | 65.33% | 65.60% | 72.08% | 56.14% |
| 2016 | 62.29% | 60.63% | 62.06% | 56.22% |
| **Total** | **64.62%** | **63.01%** | **67.50%** | **54.93%** |

For five of the eight indicators, an OA Altmetric Advantage for books is apparent: News, Blog, Policy and both Twitter coverage and Twitter rate show an AF of between 2.37 and 3.24. Wikipedia coverage shows no difference in coverage between OA books and non-OA books. Mendeley coverage is lower for OA books than non-OA books, although the average number of Mendeley readers is slightly higher.

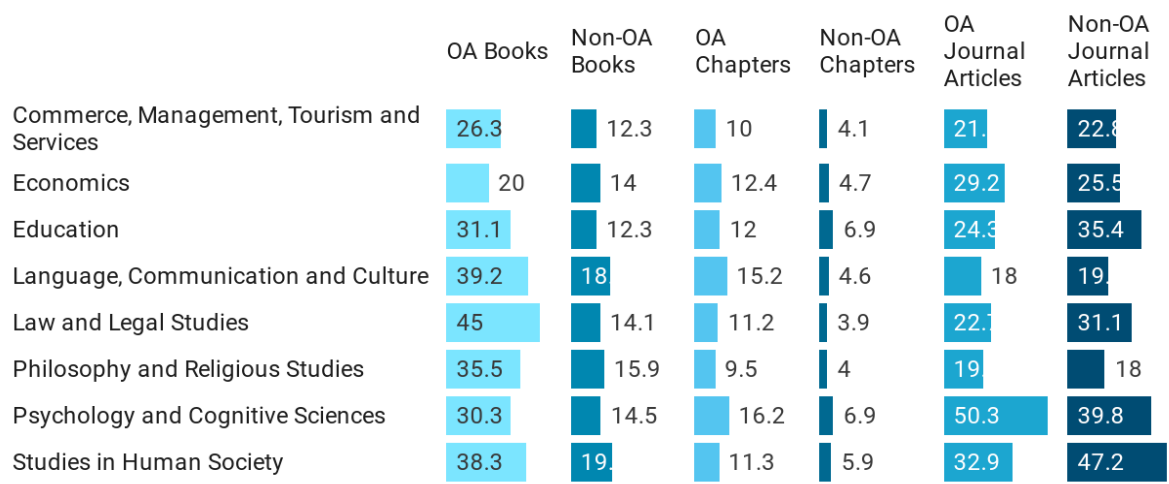

*Figure 3 Differences in Altmetric Attention Coverage Between OA and non-OA Books, Chapters and Articles Published in 2016 by Discipline*

OA books published in 2016 are more likely to get attention than non-OA books for all disciplines, with coverage falling into a similar range as that reported by journal articles published in the same disciplines and year. OA chapters get consistently more attention from Altmetric attention sources than their non-OA equivalents. The OAAA for a comparable set of journal articles is not consistently reported for all disciplines, with Altmetric Attention coverage higher for Economics and Psychology, but lower for Education and Studies in Human Society (Figure 3).

All eight indicators show a positive OA Altmetric Advantage for chapters, although generally, coverage is lower for chapters than books (Table 3). The only exception is for Mendeley coverage, where the proportion of OA chapters with Altmetric Attention is higher than OA books, and the number of Readers higher for all books and OA chapters. Other than Mendeley, Wikipedia shows the smallest AF for chapters (1.74), and News shows the highest (4.11).

*Table 3 Differences in Attention Sources Between OA and non-OA Books and Chapters*

| Discipline | Books | | | Chapters | | |
|---|---|---|---|---|---|---|
| | OA | non-OA | OAAA AF | OA | non-OA | OAAA AF |
| News coverage | 4.57%* | 1.67%* | 2.53 | 1.47%* | 0.27%* | 4.11 |
| Blog coverage | 6.66%* | 2.43%* | 2.42 | 0.68%* | 0.15%* | 3.08 |
| Wikipedia coverage | 5.22% | 4.79% | 1.09 | 0.34%* | 0.19%* | 1.74 |
| Policy coverage | 2.15%* | 0.59%* | 3.24 | 0.45%* | 0.09%* | 4.00 |
| Twitter coverage | 22.78%* | 8.94%* | 2.37 | 7.86%* | 3.39%* | 2.12 |
| Unique Twitter accounts (geometric mean) | 0.45+ | 0.13+ | 3.05 | 0.10+ | 0.04+ | 2.41 |
| Mendeley coverage | 52.42%* | 57.63%* | 0.91 | 65.75%* | 54.10%* | 1.20 |
| Mendeley readers (geometric mean) | 1.97 | 1.88 | 1.04 | 2.93+ | 1.49+ | 1.86 |

\* Significant at 5.00% using Fisher Exact Test

+ Significant at 5.00% using Student T-Test Unpaired

## Twitter

Field differences in Twitter coverage and the average number of Twitter accounts linking to the books and chapters are apparent in the data. In the data presented in Table 4a, 'Language, Communications and Culture' and 'Law and Legal Studies' have the highest Twitter coverage for books, both for OA books and non-OA books, although a clear OAAA is shown. For chapters, the highest coverage is for 'Psychology and Cognitive Sciences' and Education, both for OA chapters and non-OC chapters. In general, an OAAA AF of 2 applies to chapters. Both OA books and non-OA books get substantially higher rates of Twitter attention than their chapter equivalents.

The average number of Tweets is also higher for OA books and OA chapters (Table 4b). Although the rate is low – with a geometric mean of > 1.00 for all cohorts, OA books received higher rates of attention on Twitter than non-OA books. OA chapters have a similar average to non-OA books, and show an OAAA AF of over 2. Coverage and average Tweets appear related, with peaks corresponding across the disciplines.

There are some disciplinary differences between books and chapters: Law and 'Language, Communication and Culture' have the highest rates of Twitter coverage for books, but the same subjects are amongst the lowest for chapters, for both coverage and geometric mean.

*Table 4a Percentages of books and chapters with non-zero Altmetric.com Twitter activity*

| Discipline | Unique Twitter Accounts (Coverage, %) | | | | | |
| --- | --- | --- | --- | --- | --- | --- |
| | Books | | | Chapters | | |
| | OA | non-OA | OAAA AF | OA | non-OA | OAAA AF |
| Commerce, Management, Tourism and Services | 16.87%* | 6.25%* | 2.48 | 5.36%* | 2.27%* | 2.15 |
| Economics | 16.42%* | 9.35%* | 1.67 | 6.94%* | 2.47%* | 2.36 |
| Education | 16.04%* | 6.91%* | 2.10 | 9.95%* | 4.69%* | 1.97 |
| Language, Communication and Culture | 34.42%* | 9.73%* | 3.19 | 8.3%* | 2.90%* | 2.55 |
| Law and Legal Studies | 27.84%* | 7.72%* | 3.23 | 5.08%* | 2.28%* | 2.06 |
| Philosophy and Religious Studies | 18.18%* | 8.09%* | 2.16 | 7.06%* | 2.82%* | 2.33 |
| Psychology and Cognitive Sciences | 10.49% | 6.45% | 1.58 | 10.55%* | 4.78%* | 2.02 |
| Studies in Human Society | 23.81%* | 10.54%* | 2.14 | 7.61%* | 3.59%* | 1.97 |

\* Significant at 5.00% using Fisher Exact Test

*Table 4b Geometric mean Twitter activity of books and chapters*

| Discipline | Tweets (Geometric Mean) | | | | | |
| --- | --- | --- | --- | --- | --- | --- |
| | Books | | | Chapters | | |
| | OA | non-OA | OAAA AF | OA | non-OA | OAAA AF |
| Commerce, Management, Tourism and Services | 0.38+ | 0.09+ | 3.78 | 0.09+ | 0.02+ | 2.62 |
| Economics | 0.29+ | 0.15+ | 1.78 | 0.09+ | 0.03+ | 2.64 |
| Education | 0.31+ | 0.12+ | 2.39 | 0.12+ | 0.05+ | 2.26 |
| Language, Communication and Culture | 0.78+ | 0.13+ | 5.04 | 0.09+ | 0.03+ | 2.49 |
| Law and Legal Studies | 0.42+ | 0.11+ | 3.36 | 0.06+ | 0.02+ | 2.43 |
| Philosophy and Religious | 0.30+ | 0.11+ | 2.65 | 0.09+ | 0.03+ | 2.71 |
| Studies | | | | | | |
| Psychology and Cognitive Sciences | 0.18 | 0.10 | 1.76 | 0.13+ | 0.05+ | 2.32 |
| Studies in Human Society | 0.49+ | 0.17+ | 2.67 | 0.10+ | 0.04+ | 2.28 |

+ Significant at 5.00% using Student T-Test Unpaired

## Mendeley

Mendeley coverage of OA books is consistently lower than non-OA books for all disciplines, however Mendeley coverage of OA chapters is consistently higher. Coverage for books (both OA and non-OA) shows a low rate of variation, with the highest coverage demonstrated by Education non-OA books (64.26%), and the lowest by Economics OA books (47.01%). More variation is shown by chapter coverage. The lowest rates of coverage are shown by 'Law and Legal Studies' non-OA chapters (41.50%), the highest coverage is nearly twice as much: 'Psychology and Cognitive Sciences OA chapters (80.76%).

*Table 5a Percentages of books and chapters with non-zero Mendeley activity*

| Discipline | Mendeley Readers (Coverage, %) | | | | | |
|---|---|---|---|---|---|---|
| | Books | | | Chapters | | |
| | OA | non-OA | OAAA AF | OA | non-OA | OAAA AF |
| Commerce, Management, Tourism and Services | 53.01% | 62.43% | 0.86 | 67.79%* | 56.15%* | 1.19 |
| Economics | 47.01%* | 57.30%* | 0.83 | 56.72%* | 47.69%* | 1.17 |
| Education | 57.75% | 64.26% | 0.91 | 73.82%* | 69.61%* | 1.06 |
| Language, Communication and Culture | 50.15% | 53.19% | 0.95 | 68.75%* | 47.50%* | 1.41 |
| Law and Legal Studies | 53.61% | 55.28% | 0.97 | 53.88%* | 41.50%* | 1.27 |
| Philosophy and Religious Studies | 50.91% | 53.98% | 0.94 | 63.65%* | 43.31%* | 1.44 |
| Psychology and Cognitive Sciences | 54.55% | 60.23% | 0.91 | 80.76%* | 68.76%* | 1.16 |
| Studies in Human Society | 52.83%* | 59.87%* | 0.89 | 58.95%* | 50.46%* | 1.16 |

* Significant at 5.00% using Fisher Exact Test

The geometric mean for Mendeley readers for books shows no difference between OA and non-OA, however, an OAAA for OA chapters is shown, with a typical AF of 2. A low level of discipline variation is shown for books, with a much higher variation for chapters. 'Law and Legal Studies' and 'Philosophy and Religious Studies' have an average below two for books and chapters of all types. 'Psychology and Cognitive Sciences' shows an average Mendeley readership for 2.88 for non-OA chapters and 5.44 for OA chapters.

*Table 5b Geometric mean Mendeley activity of books and chapters*

| Discipline | Mendeley Readers (Geometric Mean) | | | | | |
|---|---|---|---|---|---|---|
| | Books | | | Chapters | | |
| | OA | non-OA | OAAA AF | OA | non-OA | OAAA AF |
| Commerce, Management, Tourism and Services | 2.75 | 2.45 | 1.12 | 3.53+ | 1.65+ | 2.01 |
| Economics | 1.96 | 1.97 | 1.00 | 2.50+ | 1.19+ | 1.92 |
| Education | 2.25 | 2.38 | 0.95 | 3.67+ | 2.26+ | 1.57 |
| Language, Communication and Culture | 1.42 | 1.57 | 0.91 | 2.54+ | 1.05+ | 2.27 |
| Law and Legal Studies | 1.89 | 1.46 | 1.28 | 1.38+ | 0.84+ | 1.58 |
| Philosophy and Religious Studies | 1.61 | 1.56 | 1.03 | 1.90+ | 0.88+ | 2.06 |
| Psychology and Cognitive Sciences | 2.77 | 2.49 | 1.11 | 5.44+ | 2.88+ | 1.80 |
| Studies in Human Society | 2.09 | 1.99 | 1.05 | 2.24+ | 1.23+ | 1.74 |

+ Significant at 5.00% using Student T-Test Unpaired

## Other altmetrics

The non-zero proportions for News, Blogs, Wikipedia and Policy Documents are low, at around or below 2% for all indicators. Although the coverage is at a low level, a clear OAAA Attention Factor for chapters is shown for News, Blogs and Policy Documents. Wikipedia is the exception, where a low to moderate OAAA AF is shown for both chapters and books. Books show higher rates of coverage than chapters for the majority of indicators.

Since the number of publications with attention from these four attention sources are much lower than Mendeley and Twitter, the Attention Factor is much more prone to being skewed by exceptions.

News coverage (Table 6a) varies greatly by discipline, with attention for all chapters and non-OA books being around 1%. In general, Psychology and 'Studies in Human Society' do better than other disciplines, for both books and chapters. A moderate OAAA for books is suggested, in contrast, OA chapters show strong OAAA, with an AF ranging from 2.94 ('Philosophy and Religious Studies') to 7.26 ('Commerce, Management Tourism and Services'). 'Psychology and Cognitive Studies' shows the highest rate for OA chapters (2.30%), non-OA chapters (0.49%).

*Table 6a Percentages of books and chapters with non-zero Altmetric.com news activity*

| Discipline | Books | | | Chapters | | |
|---|---|---|---|---|---|---|
| | OA | non-OA | OAAA | OA | non-OA | OAAA AF |

| | | | AF | | | |
|---|---|---|---|---|---|---|
| Commerce, Management, Tourism and Services | 2.41% | 1.17% | 1.95 | 1.86%* | 0.13%* | 7.26 |
| Economics | 3.73% | 1.41% | 2.37 | 1.27%* | 0.19%* | 4.22 |
| Education | 2.14% | 0.90% | 2.14 | 0.94%* | 0.20%* | 3.77 |
| Language, Communication and Culture | 4.75%* | 1.57%* | 2.78 | 1.08%* | 0.20%* | 4.18 |
| Law and Legal Studies | 5.15%* | 1.94%* | 2.47 | 1.00%* | 0.17%* | 4.46 |
| Philosophy and Religious Studies | 0.91% | 1.07% | 0.85 | 0.53%* | 0.16%* | 2.94 |
| Psychology and Cognitive Sciences | 3.50% | 1.43% | 2.29 | 2.30%* | 0.49%* | 3.69 |
| Studies in Human Society | 7.26%* | 2.27%* | 2.90 | 1.44%* | 0.33%* | 3.52 |

* Significant at 5.00% using Fisher Exact Test

Blog coverage (Table 6b) also shows disciplinary differences, although different from News coverage. The highest rate of book coverage is shown by 'Law and Legal Studies', for OA books (12.37%), 'Studies in Human Society' OB (9.75%) and Economics CB (4.85%). The highest OAAA AFs for books are shown by 'Law and Legal Studies' (4.14) and 'Language, Communication and Culture' (3.52). Blogging coverage for chapters is lower, corresponding with News coverage, being at or around 1%. Nevertheless, with the exception of 'Philosophy and Religious Studies', the OAAA AF appears to be considerable, falling between 2.82 ('Studies in Human Society') and 4.71 (Education).

*Table 6b Percentages of books and chapters with non-zero Altmetric.com blog activity*

| Discipline | Books | | | chapters | | |
|---|---|---|---|---|---|---|
| | Oa_all | non-OA | OAAA AF | Oa_all | non-OA | OAAA AF |
| Commerce, Management, Tourism and Services | 3.91% | 1.24% | 2.66 | 0.28%* | 0.06%* | 3.79 |
| Economics | 6.72% | 4.85% | 1.35 | 0.88%* | 0.15%* | 3.87 |
| Education | 3.21%* | 1.27%* | 2.26 | 0.94%* | 0.14%* | 4.71 |
| Language, Communication and Culture | 6.23%* | 1.57%* | 3.52 | 1.14%* | 0.20%* | 4.35 |
| Law and Legal Studies | 12.37%* | 2.56%* | 4.14 | 0.90%* | 0.19%* | 3.79 |
| Philosophy and Religious Studies | 2.73% | 1.83% | 1.47 | 0.13% | 0.19% | 0.71 |
| Psychology and Cognitive Sciences | 3.50% | 1.89% | 1.78 | 0.50%* | 0.07%* | 4.69 |
| Studies in Human Society | 9.75%* | 3.51%* | 2.57 | 0.60%* | 0.18%* | 2.82 |

* Significant at 5.00% using Fisher Exact Test

Wikipedia coverage (Table 6c) typically shows no OAAA for OA books, but does show an OAAA for OA chapters. Books in the disciplines of 'Language, Communication and Culture', and 'Philosophy and Religious Studies' show high rates of coverage on Wikipedia, for both OA books and non-OA books. Both show a marginal OAAA. Chapter coverage on Wikipedia is much lower, with all cohorts showing coverage of less than 1%.

*Table 6c Percentages of books and chapters with non-zero Altmetric.com Wikipedia activity*

| Discipline | Books | | | Chapters | | |
| --- | --- | --- | --- | --- | --- | --- |
| | OA | non-OA | OAAA AF | OA | non-OA | OAAA AF |
| Commerce, Management, Tourism and Services | 3.61% | 2.28% | 1.54 | 0.17% | 0.06% | 2.56 |
| Economics | 1.49% | 1.46% | 1.02 | 0.32%* | 0.09%* | 2.79 |
| Education | 1.07% | 0.99% | 1.07 | 0.31% | 0.10% | 2.69 |
| Language, Communication and Culture | 9.79% | 7.15% | 1.35 | 0.24% | 0.25% | 0.97 |
| Law and Legal Studies | 5.15% | 3.46% | 1.46 | 0.30% | 0.09% | 2.85 |
| Philosophy and Religious Studies | 9.09% | 6.43% | 1.40 | 0.27% | 0.28% | 0.97 |
| Psychology and Cognitive Sciences | 1.40% | 1.86% | 0.76 | 0.60% | 0.32% | 1.75 |
| Studies in Human Society | 5.22% | 5.52% | 0.95 | 0.33%* | 0.19%* | 1.64 |

\* Significant at 5.00% using Fisher Exact Test

Policy coverage of books shows significant disciplinary differences. Three fields ('Philosophy and Religious Education', 'Psychology and Cognitive Studies' and 'Language, Communication and Culture') receive negligible amounts of attention. Economics books (both OA and non-OA), and OA books in 'Law and Legal Studies' and 'Commerce, Management, Tourism and Services' all have around 5% coverage. Although chapters generally receive less than 1% coverage – for both OA and non-OA – OA chapters universally receive higher rates of coverage than non-OA.

*Table 6d Percentages of books and chapters with non-zero Altmetric.com policy activity*

| Discipline | Books | | | Chapters | | |
| --- | --- | --- | --- | --- | --- | --- |
| | OA | non-OA | OAAA AF | OA | non-OA | OAAA AF |
| Commerce, Management, Tourism and Services | 4.82%* | 0.78%* | 4.88 | 0.62%* | 0.08%* | 5.18 |
| Economics | 5.22% | 3.04% | 1.63 | 1.10%* | 0.21%* | 3.59 |
| Education | 2.14% | 0.68% | 2.70 | 0.47%* | 0.08%* | 4.47 |
| Language, Communication and Culture | 0.30% | 0.12% | 2.35 | 0.06% | 0.03% | 1.81 |
| Law and Legal Studies | 4.12%* | 1.23%* | 3.04 | 0.40%* | 0.04%* | 6.07 |
| Philosophy and Religious Studies | 0% | 0.06% | 0 | 0.27%* | 0.04%* | 5.68 |
| Psychology and Cognitive Sciences | 0.70% | 0.16% | 3.74 | 0.10% | 0.04% | 2.46 |
| Studies in Human Society | 2.72%* | 0.63%* | 3.76 | 0.38%* | 0.12%* | 2.75 |

\* Significant at 5.00% using Fisher Exact Test

# Discussion

An important limitation of this paper is that it only examins the OAAA across a range of disciplines within the Humanities and Social Sciences; and that in general, the number of OA books is low, forming a small percentage of the overall population of books and chapters sampled. Furthermore, the nature of the Dimensions database means that only books and chapters with DOIs are included. There are implications, therefore, for the test corpus, in that approximately half of academic books have DOIs, and these may be disproportionately derived from large commercial publishers (Grimme et al., 2019), and thus unrepresent small presses.

In general, the proportion of OA books and chapters is much smaller than the corresponding set of journal articles. In contrast with journal articles, where a clear tendency towards increasing rates of OA is confirmed, books and chapters show no such clear or sustained progression. In general, however, when ranking the disciplines by the proportion of OA output, that ranked order is consistent between journal article, and books and chapters: with Philosophy being lowest for both, and Psychology and Economics being highest for both. This suggests that the cultural preferences and practices towards OA are shared between book and journal publishing. The two highest fields have well-established archiving practises, with Psychology often being archived alongside life science and medical science materials, and with Economics having RePEc.org, a dedicated repository for economics and related sciences.

The proportions of the different types of OA for books and chapters vary, when compared against OA types for journal articles. The most frequent OA classes for both books and chapters are Bronze (implying that they are being made freely available from the publishers' websites without an explicity license) and 'Green, Submitted' (Figure 2), meaning that they have been archived on a repository, having been accepted by a publisher. Gold is the third most populous route for OA books and Hybrid being the third most populous route for chapters[5].

## Trends in the altmetrics of books and chapters

Not all altmetric attention sources have been examined: with lower-frequency indicators (e.g. Facebook, Reddit) being discarded, as well as sources that weren't being captured by Altmetric throughout 2013-2016 (e.g. Patents, Sina Weibo).

---

5  https://support-funder.dimensions.ai/support/solutions/articles/13000042712-where-does-the-definition-of-open-access-come-from-in-dimensions-what-does-it-include-

The figures for overall coverage in Figure 3 may be contrasted with previous findings that reported Altmetric coverage for journal articles published in 2011 and 2013 as rising from 10.8% of publications in 2011, 23.8% in 2012 and over 25% in 2013, (Costas, Zahedi, & Wouters, 2014).

Care needs to be taken in the interpretation of this data:

1. The number of OA books presented in this study is low, and at a discipline level, differences are not uniformly found to be significant.

2. The books and chapters analysed in this paper - and the journal articles shown in Figure 3 - have had between 3 and 6 years to accrue attention. In contrast, the journal articles analysed by Costas et al and Taylor had had 2-4 years and 1 year, respectively.

3. All of the books and chapters in this study have been registered with Crossref and have a DOI, and therefore previous suggestions that the presence of a DOI might be sufficient to explain higher rates of activity for OA publications are ruled out (Calver & Bradley, 2009).

The relative difference in population sizes suggests some difficulties in making like-for-like comparisons. Nevertheless, this data suggests evidence for an OAAA for OA books and chapters across several of altmetric indicators, and this is generally confirmed at a 5% signficance using the Fisher Exact test, which is well-optimized for differences in population size.

The proportion of OA books getting attention is higher on Twitter (by a factor of 2.4), News (2.5), Blogs (2.4) and Policies (3.2). OA books also get attention from more Twitter accounts (3.1). The proportion of OA chapters getting attention on Twitter is higher (by a factor of 2.1), News (4.1), Blogs (3.1), Wikipedia (1.7) and Policies (4.0). OA chapters get attention from more Twitter accounts (2.4) and more Mendeley readers (1.9).

No OAAA is found for OA books on Wikipedia, proportion of OA books and chapters on Mendeley, and average Mendeley readership for OA books.

In general, therefore, we feel confident in concluding that the OAAA that has been observed for journal articles is generally shared by both books and chapters.

The magnitude of OAAA for both books and chapters varies across the disciplines studied, for both the overall proportion of altmetric attention, and by attention source.

Zahedi, Costas, & Wouters, (2014) have previously reported substantial field differences for Twitter, 'Social and Behavioural Sciences' receiving considerably more attention for journal articles than Language, Law, Arts & Humanities. These observations are not found for OA books, suggesting that the OA status of books in these areas significantly increases the propability of sharing on social networks.

In general, the OAAA for all disciplines for Mendeley is either non-existent or relatively small, suggesting that academic users of books are relatively unaffected by OA status at a book level, although a general chapter-level OAAA is observed.

Similar phenomena are reported for Wikipedia coverage, with book-level citations largely unaffected by OA status; and a significant range of OAAAs found at a chapter level. The potential importance of exogenous agents that may affect Wikipedia coverage – for example, Oabot[6], that creates links to OA articles - has been discussed in recent research looking at journal articles (Holmberg, Hedman, et al., 2019). However there is no evidence to suggest that this software either leads Wikipedia editors to preferentially use OA materials, or for the citations to be preferentially discovered by altmetric suppliers. Indeed, the similar to the trends observed by Mendeley – which is largely used by academics – suggests a shared lack of importance in OA status at a book level, but a marginally raised use at a chapter level.

The disciplines examined here show marked disciplinary differences, extending earlier observations of the altmetrics of journal articles. This finding emphasizes the importance of either either normalizing for discipline, or taking care to only compare sets of documents on a like-for-like basis. This should be extended to include normalizing for publication type: chapters, books, and journal articles show different trends, and their expected performance may not be easily extrapolated.

While the absent, or reduced OAAA for the two attention sources that might be considered to be closer to the academic ecosystem (Mendeley and Wikipedia) reinforces the mixed results found by earlier research for journals, and suggests that OA status is not considered to be an important factor in choosing academic material. However, the significant OAAA shown by the more broadly used and authored attention sources, suggest that OA status has a significant affect on this broader impact, and as such, indicates that OA policies are broadly successful in expanding the impact of research.

# Conclusions

This research does not attempt to investigate the underlying causes of the OAAA, indeed there has been little causative analysis on the underlying mechanisms of OA for journals, with a general assumption being made that 'more access allows more people to read' (Piwowar et al., 2018).

Craig, Plume, McVeigh, Pringle, & Amin (2007) summarized three possible underlying mechanisms to explain the OACA:

1. That more access allows more people to read (the OA postulate),
2. That authors choose to make their best work available freely (the selection bias postulate),
3. That OA articles get attention earlier (the early view postulate).

Fourth and fifth possible postulates would be that:

1. The growth in OA is disproportionately growing scholarship (an output inflation postulate),
2. That increasing rates of citation or sharing behaviour (a usage inflation postulate).

---

6   https://en.wikipedia.org/wiki/Wikipedia:OABOT

Early research (Moed, 2007) concluded that postulates 2 and 3 were likely to be the mechanisms by which the OACA was effected, with there being no clear evidence of a general OA effect.

There should be no assumption that any postulate that is true for citation behaviour should be the same for that measured by altmetrics, or that altmetrics share a common underlying cause. Although there has been research into why people believe some outputs are shared disproportionately (Holmberg & Vainio, 2018) and intention use of Mendeley (Mohammadi, Thelwall, Haustein, & Larivière, 2015), in general, this area remains unexplored.

However, the growth of OA publishing has provoked some research into the nature of the published corpus, and how it might change over time. As OA has developed over the subsequent years, the publishing environment has evolved to accommodate the new business model, with both author, institutional and reader behavioural changes. The growth of the OA mega-journal has enabled growth in publishing 'technically sound science without consideration of novelty' (Brainard, 2019), journals that are likely to be cited at a lower rate than specialized journals (Piwowar et al., 2018b). Institutions and funders (Berg, 2010) are increasingly mandating OA, however the actual rate of making research available via repositories – even when there are no publisher restrictions – may remain low (Koos, 2019).

The behaviour of OA books and chapters sheds some interesting light on the potential underlying mechanisms of an OAAA: and that this is particularly important as both evidence for an OACA or an OAAA may be used as evidence to increase rates of OA publishing for books and chapters. Future work could additionally focus on the mechanism by which chapters accrue attention, and whether this attention adds to book level attention, or whether it substitutes it.

Interpretation of these varied results supports us to explore how the different postulates could be applying for different attention sources and fields of study. Research in the mathematics of citation and sharing has suggested a two-stage (Didegah & Thelwall, 2013) – or binomial – process, and it is possible that the data presented in this research supports this concept. The most populous indicator (Mendeley) is highly integrated into scholarly research workflows, and is largely pre-populated with metadata from both users and Scopus: the selection process is likely to be driven by the appropriateness of the researcher to a researchers' search terms, with the openness-or-not of the research likely to be a secondary criteria.

In contrast, the other attention sources presented here are not integrated into the infrastructure of scholarship: research needs to be introduced into these networks in order to be shared, and it is here where either infrastructure (e.g., Twitter or Wikipedia bots), agency (e.g., researchers promoting publications to bloggers), or existing practise (e.g., widespread adoption of subject repositories such as Repec) are likely to play a role in bringing people's attention to research, with issues of relevance and quality being a secondary stage.

For books, the rate of Gold OA publishing is very low, with Green, Submitted being the dominant route to OA, so the key drivers behind the access of books would appear to be whether an author chooses to make their outputs available via repositories, and whether a publisher permits this action. It is not unreasonable to suggest that there is, therefore, a sense of agency behind the selection that supports an author to act. This may, itself, have a number of components: an author may be obliged by an institution or funder; an author may believe this book to be of a particularly high quality, and therefore more deserving of a wider audience. Finally, an author may be a student of bibliometrics, and conclude that the best way to optimize citation rates is to make a book available on repositories. However, the principal route to OA for chapters is Gold or Bronze (in other words, the publisher makes the content available freely from their website), and the OAAA is still true.

Therefore, with Blogs and News attention, it is possible that there are two phenomena at play. Firstly, an author or publisher might choose to promote an OA book or chapter in a way that they wouldn't if it were non-OA, and investing more effort into the promotion. Secondly, if quality was the driver for the author to make the chapter or book OA, it is possible that this is the cause of making it more likely to be discussed by News or Blog sources. In the case of Policy attention, which is likely to accrue much slower than any other source in this study, the issue of quality is more likely to come to the forefront.

Stepping back from the well-documented OACA and examining the complexities of the OAAA gives us new insights into understanding the potential complexities of behaviour and access, and how these may shift over time as the various stakeholders in the community adapt their performance. The complexities suggest firstly, that there is no simple, 'fixed view' of any OA advantage, and that behaviour and performance needs to be periodically benchmarked in order to understand objective academic performance: and that this is all the more important if this citation and sharing performance are being taken into account by the stakeholders involved in moving the research ecosystem towards a world more dominated by OA research.

Secondly, the lack of research into the underlying mechanisms that produce the OACA and OAAA implies that these decisions are, effectively, 'black boxes', where the only observations are the inputs and the outputs. Additional research into the mechanisms and causes are required.

The observation that the OAAA exists for both books and chapters, despite having largely different routes to OA when compared to journal articles, suggests that all routes are successful in boosting attention and impact, as measured through altmetrics; and that advocates of OA books and chapters could prefer *either* the Gold/Bronze route *or* the self-archiving Green route and expect to see increased rates of social use, share and impact.

# References

Adie, E. (2014). *Attention! A study of open access vs non-open access articles*. https://doi.org/10.6084/M9.FIGSHARE.1213690.V1

Almind, T. C., & Ingwersen, P. (1997). Informetric analyses on the world wide web: methodological approaches to 'webometrics.' *Journal of Documentation*, *53*(4), 404–426. https://doi.org/10.1108/EUM0000000007205

Altmetric. (2018). Patent data in Altmetric highlights the commercialization of research – Altmetric. Retrieved December 16, 2019, from https://www.altmetric.com/press/press-releases/patent-data-in-altmetric-highlights-the-commercialization-of-research/

Bar-Ilan, J. (2000). The web as an information source on informetrics? A content analysis. *Journal of the American Society for Information Science and Technology*, *51*(5), 432–443. https://doi.org/10.1002/(sici)1097-4571(2000)51:5<432::aid-asi4>3.0.co;2-7

Berg, J. (NIH). (2010). Measuring the Scientific Output and Impact of NIGMS Grants – NIGMS Feedback Loop Blog – National Institute of General Medical Sciences. Retrieved December 26, 2019, from https://loop.nigms.nih.gov/2010/09/measuring-the-scientific-output-and-impact-of-nigms-grants/

Björk, B.-C., Welling, P., Laakso, M., Majlender, P., Hedlund, T., & Guðnason, G. (2010). Open Access to the Scientific Journal Literature: Situation 2009. *PLoS ONE*, *5*(6), e11273. https://doi.org/10.1371/journal.pone.0011273

Bode, C., Herzog, C., Hood, D., & McGrath, R. (2019). *A Guide to the Dimensions Data Approach*. https://doi.org/10.6084/M9.FIGSHARE.5783094.V7

Bornmann, L. (2013). What is societal impact of research and how can it be assessed? a literature survey. *Journal of the American Society for Information Science and Technology*, *64*(2), 217–233. https://doi.org/10.1002/asi.22803

Bornmann, L. (2014). Validity of altmetrics data for measuring societal impact: A study using data from Altmetric and F1000Prime. *Journal of Informetrics*, *8*(4), 935–950. https://doi.org/10.1016/j.joi.2014.09.007

Brainard, J. (2019, September 13). Open-access megajournals lose momentum. *Science*, Vol. 365, p. 1067. https://doi.org/10.1126/science.365.6458.1067

Britt Holbrook, J., & Frodeman, R. (2011). Peer review and the ex ante assessment of societal impacts. *Research Evaluation*, *20*(3), 239–246. https://doi.org/10.3152/095820211X12941371876788

Calver, M. C., & Bradley, J. S. (2009). Patterns of Citations of Open Access and Non-Open Access Conservation Biology Journal Papers and Book Chapters. *Conservation Biology*, *24*(3), 872–880. https://doi.org/10.1111/j.1523-1739.2010.01509.x

Chi, P. S. (2016). Differing disciplinary citation concentration patterns of book and journal literature? *Journal of Informetrics*. https://doi.org/10.1016/j.joi.2016.05.005

Clarivate. (2020). Book Citation Index - Clarivate Analytics. Retrieved January 17, 2020, from http://wokinfo.com/products_tools/multidisciplinary/bookcitationindex/

Commision, E. (n.d.). Open Access to scientific information | Digital Single Market. Retrieved January 17, 2020, from https://ec.europa.eu/digital-single-market/en/policies/open-access

Costas, R., Zahedi, Z., & Wouters, P. (2014). Do "altmetrics" correlate with citations? Extensive comparison of altmetric indicators with citations from a multidisciplinary perspective. *Journal of the Association for Information Science and Technology*, *66*(10), n/a-n/a. https://doi.org/10.1002/asi.23309

Craig, I., Plume, A., McVeigh, M., Pringle, J., & Amin, M. (2007). Do open access articles have greater citation impact?A critical review of the literature. *Journal of Informetrics*, *1*(3), 239–248. https://doi.org/10.1016/j.joi.2007.04.001

Davis, P. M., Lewenstein, B. V, Simon, D. H., Booth, J. G., & Connolly, M. J. L. (2008). Open access publishing, article downloads, and citations: randomised controlled trial. *BMJ*, *337*(jul31 1), a568–a568. https://doi.org/10.1136/bmj.a568

Didegah, F., Ghaseminik, Z., & Alperin, J. P. (2018). *Using a diabetes discussion forum and Wikipedia to detect the alignment of public interests and the research literature Background Methodology / Principal findings Conclusions / Significance*.

Didegah, F., & Thelwall, M. (2013). Which factors help authors produce the highest impact research? Collaboration, journal and document properties. *Journal of Informetrics*, *7*(4), 861–873. https://doi.org/10.1016/j.joi.2013.08.006

Elsevier. (2020a). Books | Elsevier Scopus Blog. Retrieved January 17, 2020, from https://blog.scopus.com/topics/books

Elsevier. (2020b). Open Access Books. Retrieved January 17, 2020, from https://www.elsevier.com/about/open-science/open-access/open-access-books

Eysenbach, G. (2011). Can tweets predict citations? Metrics of social impact based on Twitter and correlation with traditional metrics of scientific impact. *Journal of Medical Internet Research*. https://doi.org/10.2196/jmir.2012

Frantsvåg, J. E., & Strømme, T. E. (2019). Few Open Access Journals Are Compliant with Plan S. *Publications*, *7*(2), 26. https://doi.org/10.3390/publications7020026

Grimme, S., Taylor, M., Elliott, M. A., Holland, C., Potter, P., & Watkinson, C. (2019). *The State of Open Monographs*. https://doi.org/10.6084/M9.FIGSHARE.8197625.V4

Halevi, G., Nicolas, B., & Bar-Ilan, J. (2016). The Complexity of Measuring the Impact of Books. *Publishing Research Quarterly*, *32*(3), 187–200. https://doi.org/10.1007/s12109-016-9464-5

Hammarfelt, B. (2014). Using altmetrics for assessing research impact in the humanities. *Scientometrics*, *101*(2), 1419–1430. https://doi.org/10.1007/s11192-014-1261-3

Hawkins, D. T. (2016). Altmetrics and Books: Bookmetrix and Other Implementations. *Against the Grain*, *28*(3). https://doi.org/10.7771/2380-176x.7364

Health, N. I. of. (n.d.). NIH Public Access Policy Details | publicaccess.nih.gov. Retrieved January 17, 2020, from https://publicaccess.nih.gov/policy.htm

Heilman, J. M., Kemmann, E., Bonert, M., Chatterjee, A., Ragar, B., Beards, G. M., … Laurent, M. R. (2011). Wikipedia: A Key Tool for Global Public Health Promotion. *Journal of Medical Internet Research*, *13*(1), e14. https://doi.org/10.2196/jmir.1589

Holbrook, J. B. (2019). *Philosopher's Corner: Open Science, Open Access, and the Democratization of Knowledge*. 26–28. Retrieved from https://issues.org/philosophers-corner-open-science-open-access-and-the-democratization-of-knowledge/

Holmberg, K., Bowman, S., Bowman, T., Didegah, F., & Kortelainen, T. (2019). What Is Societal Impact and Where Do Altmetrics Fit into the Equation? *Journal of Altmetrics*, *2*(1). https://doi.org/10.29024/joa.21

Holmberg, K., Hedman, J., Bowman, T. D., Didegah, F., & Laakso, M. (2019). Do articles in open access journals have more frequent altmetric activity than articles in subscription-based journals? An investigation of the research output of Finnish universities. *Scientometrics*. https://doi.org/10.1007/s11192-019-03301-x

Holmberg, K., & Thelwall, M. (2014). Disciplinary differences in Twitter scholarly communication. *Scientometrics*, 1–16. https://doi.org/10.1007/s11192-014-1229-3

Holmberg, K., & Vainio, J. (2018). Why do some research articles receive more online attention and higher altmetrics? Reasons for online success according to the authors. *Scientometrics*. https://doi.org/10.1007/s11192-018-2710-1

Hook, D. W., Porter, S. J., & Herzog, C. (2018). Dimensions: Building Context for Search and Evaluation. *Frontiers in Research Metrics and Analytics*, *3*. https://doi.org/10.3389/frma.2018.00023

Koos, J. A. (2019). Green Deposit Rates in LIS Taylor & Francis Journals: Are Librarians "Practicing What They Preach?" *Evidence Based Library and Information Practice*, *14*(2), 137–139. https://doi.org/10.18438/eblip29560

Kousha, K., & Thelwall, M. (2015). Web indicators for research evaluation. Part 3: books and non standard outputs. *El Profesional de La Información*, *24*(6), 724. https://doi.org/10.3145/epi.2015.nov.04

Kurata, K., Morioka, T., Yokoi, K., & Matsubayashi, M. (2013). Remarkable Growth of Open Access in the Biomedical Field: Analysis of PubMed Articles from 2006 to 2010. *PLoS ONE*, *8*(5), e60925. https://doi.org/10.1371/journal.pone.0060925

Kurtz, M. J., & Henneken, E. A. (2007). *Open Access does not increase citations for research articles from The Astrophysical Journal*. Retrieved from http://arxiv.org/abs/0709.0896

McLeish, B. (Altmetric). (2016, September). Altmetric and Policy: Discovering how your research impacted real-world practises. Retrieved December 16, 2019, from Altmetric.com website: https://www.altmetric.com/blog/altmetric-and-policy-discovering-how-your-research-impacted-real-world-practises/

Moed, H. F. (2007). The effect of "open access" on citation impact: An analysis of ArXiv's condensed matter section. *Journal of the American Society for Information Science and Technology*, *58*(13), 2047–2054. https://doi.org/10.1002/asi.20663

Mohammadi, E., Thelwall, M., Haustein, S., & Larivière, V. (2015). Who reads research articles? An altmetrics analysis of Mendeley user categories. *Journal of the Association for Information Science and Technology*. https://doi.org/10.1002/asi.23286

Mohammadi, E., Thelwall, M., Kwasny, M., & Holmes, K. L. (2018). Academic information on Twitter: A user survey. *PLOS ONE*, *13*(5), e0197265. https://doi.org/10.1371/journal.pone.0197265

O'Leary, B., & Hawkins, K. (2019). *Exploring Open Access Ebook Usage*.

OAPEN. (2020). List of compliant book publishers | OAPEN. Retrieved January 17, 2020, from http://oapen.org/content/deposit-publishers-list-compliant-book-publishers

Ottaviani, J. (2016). The Post-Embargo Open Access Citation Advantage: It Exists (Probably), It's Modest (Usually), and the Rich Get Richer (of Course). *PLOS ONE*, *11*(8), e0159614. https://doi.org/10.1371/journal.pone.0159614

Phillips, D. P., Kanter, E. J., Bednarczyk, B., & Tastard, P. L. (1991). Importance of the Lay Press in the Transmission of Medical Knowledge to the Scientific Community. *The New England Journal of Medicine*.

Pinter, F., & Thatcher, S. (2012). Efficient and Effective Funding of Open Access "Books." In R. Kubilius (Ed.), *Anything Goes* (pp. 11–11). https://doi.org/10.5703/1288284314815

Piwowar, H., Priem, J., Larivière, V., Alperin, J. P., Matthias, L., Norlander, B., … Haustein, S. (2018a). The state of OA: a large-scale analysis of the prevalence and impact of Open Access articles. *PeerJ*, *6*, e4375. https://doi.org/10.7717/peerj.4375

Piwowar, H., Priem, J., Larivière, V., Alperin, J. P., Matthias, L., Norlander, B., … Haustein, S. (2018b). The state of OA: a large-scale analysis of the prevalence and impact of Open Access articles. *PeerJ*, *6*, e4375. https://doi.org/10.7717/peerj.4375

Priem, J., Taraborelli, D., Groth, P., & Neylon, C. (2010). Alt-metrics: a manifesto. Retrieved from October website: http://altmetrics.org/manifesto/

Pulido, C. M., Redondo-Sama, G., Sordé-Martí, T., & Flecha, R. (2018). Social impact in social media: A new method to evaluate the social impact of research. *PLOS ONE*, *13*(8), e0203117. https://doi.org/10.1371/journal.pone.0203117

Research, C. I. of H. (n.d.). Tri-Agency Open Access Policy on Publications - CIHR. Retrieved January 17, 2020, from https://cihr-irsc.gc.ca/e/32005.html

S, C. (n.d.). "Plan S" and "cOAlition S" – Accelerating the transition to full and immediate Open Access to scientific publications. Retrieved January 17, 2020, from https://www.coalition-s.org/

Schiltz, M. (2018). Science Without Publication Paywalls: cOAlition S for the Realisation of Full and Immediate Open Access. *Frontiers in Neuroscience*, *12*. https://doi.org/10.3389/fnins.2018.00656

Science Europe. (2019). *Briefing Paper on Open Access to Academic Books - Science Europe*. Retrieved from https://www.scienceeurope.org/our-resources/briefing-paper-on-open-access-to-academic-books/

Snijder, R. (2016). Revisiting an open access monograph experiment: measuring citations and tweets 5 years later. *Scientometrics*, *109*(3), 1855–1875. https://doi.org/10.1007/s11192-016-2160-6

Suber, P. (2012). *Open access*. Retrieved from https://mitpress.mit.edu/books/open-access

Sugimoto, C. R., & Larivière, V. (2017). Altmetrics: Broadening Impact or Amplifying Voices? *ACS Central Science*, *3*(7), 674–676. https://doi.org/10.1021/acscentsci.7b00249

Taylor, M. (2015a). *Data for "Engineers Don't Blog and Other Stories: Why Scopus Uses Subject Area Benchmarks." 1*. https://doi.org/10.17632/SMJJ59MBMB.1

Taylor, M. (2015b). *Engineers Don't Blog and Other Stories (why Scopus uses subject area benchmarking)*. https://doi.org/doi:10.6084/m9.figshare.1568135

Taylor, M. (2020). *Altmetrics and Open Access Books and Chapters*. https://doi.org/10.6084/m9.figshare.11527962

Teplitskiy, M., Lu, G., & Duede, E. (2017). Amplifying the impact of open access: Wikipedia and the diffusion of science. *Journal of the Association for Information Science and Technology*, *68*(9), 2116–2127. https://doi.org/10.1002/asi.23687

Thelwall, M. (2000). Web impact factors and search engine coverage. *Journal of Documentation*, *56*(2), 185–189. https://doi.org/10.1108/00220410010803801

Thelwall, M. (2017). Three practical field normalised alternative indicator formulae for research evaluation. *Journal of Informetrics*, *11*(1), 128–151. https://doi.org/10.1016/j.joi.2016.12.002

Thelwall, M., & Fairclough, R. (2015). Geometric journal impact factors correcting for individual highly cited articles. *Journal of Informetrics*. https://doi.org/10.1016/j.joi.2015.02.004

Thelwall, M., Haustein, S., Larivière, V., & Sugimoto, C. R. (2013). Do altmetrics work? Twitter and ten other social web services. *PloS One*, *8*(5), e64841. https://doi.org/10.1371/journal.pone.0064841

Torres-Salinas, D., Gorraiz, J., & Robinson-Garcia, N. (2018). The insoluble problems of books: what does Altmetric.com have to offer? *Aslib Journal of Information Management*, *70*(6), 691–707. https://doi.org/10.1108/AJIM-06-2018-0152

Torres-Salinas, D., Robinson-Garcia, N., & Gorraiz, J. (2017). Filling the citation gap: measuring the multidimensional impact of the academic book at institutional level with PlumX. *Scientometrics*, *113*(3), 1371–1384. https://doi.org/10.1007/s11192-017-2539-z

Umstattd, L. J., Banks, M. A., Ellis, J. I., & Dellavalle, R. P. (2008). Open Access Dermatology Publishing: No Citation Advantage Yet. *The Open Dermatology Journal*, *2*(1), 69–72. https://doi.org/10.2174/1874372200802010069

Wennström, S., Schubert, G., Stone, G., Sondervan, J., Wennström, S., Schubert, G., … Sondervan, J. (2019). *The significant difference in impact: An exploratory study about the meaning and value of metrics for open access monographs*.

Wiley. (2020). Self-Archiving | Wiley. Retrieved January 17, 2020, from https://authorservices.wiley.com/author-resources/Journal-Authors/licensing/self-archiving.html

Williams, K. (2018). Three strategies for attaining legitimacy in policy knowledge: Coherence in identity, process and outcome. *Public Administration*, *96*(1), 53–69. https://doi.org/10.1111/padm.12385

Zahedi, Z., Costas, R., & Wouters, P. (2014). How well developed are altmetrics? A cross-disciplinary analysis of the presence of 'alternative metrics' in scientific publications. *Scientometrics*. https://doi.org/10.1007/s11192-014-1264-0

# Appendix

*Appendix Table 1 Volume of Books and Chapters by Discipline (Dimensions, retrieved October 22, 2019)*

| Discipline | Books | Chapters |
|---|---|---|
| Commerce, Management, Tourism and Services | 1619 | 24,200 |
| Economics | 1909 | 27,132 |
| Education | 2400 | 18,049 |
| Language, Communication and Culture | 7915 | 27,073 |
| Law and Legal Studies | 2208 | 15,237 |
| Philosophy and Religious Studies | 3285 | 14,919 |
| Psychology and Cognitive Sciences | 3211 | 39,477 |
| Studies in Human Society | 9675 | 54,440 |

*Appendix Table 2 Total Open Books and Chapters Published by Discipline and Total Volume (% Open) (Dimensions, retrieved October 22, 2019)*

| Discipline | Publication Type | 2013 | 2014 | 2015 | 2016 | Total |
|---|---|---|---|---|---|---|
| **Commerce, Management, Tourism and Services** | Books | 11 (2.4%) | 24 (7.0%) | 18 (8.2%) | 30 (5.1%) | 83 (5.1%) |
| | Chapters | 297 (5.6%) | 371 (7.2%) | 421 (5.6%) | 684 (11.1%) | 1773 (7.3%) |
| | Journal Articles | | | | | |
| **Economics** | Books | 36 (5.8%) | 20 (5.6%) | 36 (10.2%) | 42 (7.3%) | 134 (7.0%) |
| | Chapters | 563 (9.0%) | 594 (9.4%) | 667 (8.8%) | 1002 (14.3%) | 2826 (10.4%) |
| | Journal Articles | | | | | |
| **Education** | Books | 55 (5.8%) | 59 (10.3%) | 36 (8.5%) | 37 (8.2%) | 187 (7.8%) |
| | Chapters | 228 (6.2%) | 238 (5.5%) | 407 (8.2%) | 403 (7.9%) | 1276 (7.1%) |
| | Journal Articles | | | | | |
| **Language, Communication and Culture** | Books | 67 (3.3%) | 69 (4.3%) | 79 (6.4%) | 122 (4.0%) | 337 (4.3%) |
| | Chapters | 354 (6.0%) | 358 (6.2%) | 434 (5.5%) | 518 (6.9%) | 1664 (6.2%) |
| | Journal Articles | | | | | |
| **Law and Legal Studies** | Books | 16 (2.4%) | 21 (5.2%) | 24 (6.9%) | 36 (4.5%) | 97 (4.4%) |
| | Chapters | 251 (6.6%) | 167 (5.2%) | 220 (5.2%) | 366 (9.2%) | 1004 (6.6%) |
| | Journal Articles | | | | | |
| **Philosophy and Religious Studies** | Books | 22 (2.7%) | 20 (2.8%) | 30 (5.7%) | 38 (3.1%) | 110 (3.4%) |
| | Chapters | 153 (4.4%) | 187 (6.3%) | 193 (5.3%) | 218 (4.5%) | 751 (5.0%) |

|  | Journal Articles |  |  |  |  |  |
| --- | --- | --- | --- | --- | --- | --- |
| **Psychology and Cognitive Sciences** | Books | 56 (4.5%) | 29 (3.1%) | 27 (5.9%) | 31 (5.4%) | 143 (4.5%) |
|  | Chapters | 722 (8.6%) | 658 (7.9%) | 773 (6.7%) | 851 (7.9%) | 3004 (7.6%) |
|  | Journal Articles |  |  |  |  |  |
| **Studies in Human Society** | Books | 116 (4.3%) | 82 (4.3%) | 99 (5.8%) | 144 (4.6%) | 441 (4.6%) |
|  | Chapters | 692 (5.5%) | 760 (6.4%) | 928 (6.2%) | 1311 (8.8%) | 3691 (6.8%) |
|  | Journal Articles |  |  |  |  |  |